\documentclass[fleqn,usenatbib]{mnras}

\usepackage{newtxtext,newtxmath}

\usepackage[T1]{fontenc}

\DeclareRobustCommand{\VAN}[3]{#2}
\let\VANthebibliography\thebibliography
\def\thebibliography{\DeclareRobustCommand{\VAN}[3]{##3}\VANthebibliography}

\usepackage{graphicx}	
\usepackage{amsmath}	

\newcommand\nhi{$\rm{N}_{\rm{HI}}$} 
\newcommand\lya{Ly$\alpha$} 
\newcommand\ha{H$\alpha$} 
\newcommand\oi{[O\textsc{i}]} 
\newcommand\hi{H\textsc{i}} 

\title[]{Tidally offset neutral gas in Lyman continuum emitting galaxy Haro 11}

\author[A. Le Reste et al.]{
Alexandra Le Reste,$^{1}$\thanks{E-mail:alexandra.lereste@astro.su.se}
John M. Cannon,$^{2}$
Matthew J. Hayes,$^{1}$
John L. Inoue,$^{2}$
Amanda A. Kepley,$^{3}$
\newauthor
Jens Melinder,$^{1}$
Veronica Menacho,$^{1}$
Angela Adamo,$^{1}$
Arjan Bik,$^{1}$
Timmy Ejdetj\"{a}rn,$^{1}$
\newauthor
Gyula I. G. J\'{o}zsa,$^{4,5}$
G\"{o}ran \"{O}stlin,$^{1}$
Sarah H. Taft,$^{2,6}$
\\
$^{1}$The Oskar Klein Centre, Department of Astronomy, Stockholm University; AlbaNova, SE- 10691 Stockholm, Sweden\\
$^{2}$Department of Physics and Astronomy, Macalester College; 1600 Grand Avenue, Saint Paul, 10 MN 55105, USA\\
$^{3}$National Radio Astronomy Observatory; 520 Edgemont Road, Charlottesville, VA 22903-2475, USA.\\
$^{4}$Max-Planck-Institut f\"{u}r Radioastronomie; Auf dem H\"{u}gel 69, D-53121 Bonn, Germany.\\
$^{5}$Department of Physics and Electronics, Rhodes University; P.O. Box 94, Makhanda, 6140, 15 South Africa.\\
$^{6}$Minnesota Institute for Astrophysics, School of Physics \& Astronomy, University of Minnesota; 116 Church St. SE, Minneapolis, MN 55455, USA.
}

\date{Accepted XXX. Received YYY; in original form ZZZ}

\pubyear{2023}

\begin{document}
\label{firstpage}
\pagerange{\pageref{firstpage}--\pageref{lastpage}}
\maketitle

\begin{abstract}
Around 400 million years after the Big Bang, the ultraviolet emission from star-forming galaxies reionized the Universe. Ionizing radiation (Lyman Continuum, LyC) is absorbed by cold neutral hydrogen gas (\hi) within galaxies, hindering the escape of LyC photons. Since the \hi\ reservoir of LyC emitters has never been mapped, major uncertainties remain on how LyC photons escape galaxies and ionize the intergalactic medium. We have directly imaged the neutral gas in the nearby reionization-era analog galaxy Haro 11 with the 21cm line to identify the mechanism enabling ionizing radiation escape. We find that merger-driven interactions have caused a bulk offset of the neutral gas by about $6\,$kpc from the center of the galaxy, where LyC emission production sites are located. This could facilitate the escape of ionizing radiation into our line of sight. Galaxy interactions can cause both elevated LyC production and large-scale displacement of \hi\ from the regions where these photons are produced. They could contribute to the anisotropic escape of LyC radiation from galaxies and the reionization of the Universe. We argue for a systematic assessment of the effect of environment on LyC production and escape.
\end{abstract}

\begin{keywords}
galaxies: ISM -- galaxies: starburst -- ISM: lines and bands -- galaxies: interactions -- radio lines: galaxies -- ultraviolet: galaxies
\end{keywords}



\section{Introduction}
The Universe underwent a major phase change during which almost all intergalactic hydrogen was ionized about 13 billion years ago: cosmic reionization \citep{Barkana2006}. Reionization is thought to have been driven by strong ultraviolet (UV) emission from massive stars within primeval star-forming galaxies \citep{Robertson2010,Kulkarni2019,Dayal2020,Trebitsch2023}. LyC emission is absorbed by cold neutral hydrogen gas in the interstellar medium of galaxies, hindering the escape of ionizing radiation to the intergalactic medium. Observations of galaxies at this epoch lack spatial resolution, thus nearby analogs of early galaxies have been used to understand the detailed physical processes responsible for reionization \citep{Izotov2016,Senchyna2017,Izotov2018,Berg2019,Flury2022a}. UV absorption line measurements suggest that the main property driving LyC escape is a low covering fraction of the neutral gas \citep{Gazagnes2018,Saldana-Lopez2022}, with LyC photons escaping through ionized channels within the interstellar medium. However, absorption measurements do not characterize the interstellar medium in the full physical volume covered by dense neutral gas in galaxies. 

The 21cm line of Hydrogen is the only direct tracer of \hi\ gas that can probe the entire extent of the material inhibiting LyC escape. \hi\ has never been mapped in confirmed LyC-emitters, leaving major uncertainties on how LyC photons escape galaxies and ionize the intergalactic medium \citep{MacHattie2014,Pardy2016,Puschnig2017,Taft2019}. Furthermore, even with upcoming state-of-the-art facilities, it will be impossible to observe resolved \hi\ in emission in galaxies at the Epoch of Reionization \citep{Ghara2017}. Thus, observing \hi\ in nearby galaxies with detected ionizing LyC emission is necessary for a complete understanding of the gas removal and ionization mechanisms at play during the Epoch of Reionization.

Haro 11 is the closest (D=93 Mpc) confirmed LyC-emitting galaxy \citep{Bergvall2000}, and one of only three known LyC-leaking galaxies that are sufficiently close to be observed and resolved in 21cm \citep{Bergvall2006,Leitet2013,Leitherer2016}. The UV photons in this star-forming galaxy are produced in three distinct regions \citep{Kunth2003,Adamo2010,Sirressi2022} labeled Knot A, B and C, shown in Fig. \ref{fig:21cm_global} panel B, with Knot B having the largest photo-ionization rate of all three regions \citep{Sirressi2022}. 
The global neutral gas distribution that can allow for the anisotropic escape of ionizing radiation out of the galaxy has not been measured. The molecular gas distribution has been measured \citep{Gao2022}, but molecular gas is a negligible source of LyC opacity in galaxies as compared to neutral gas. Previous single dish observations of Haro 11 detected the 21cm in emission \citep{Pardy2016}. However, the distribution of the gas has remained unknown due to insufficient sensitivity of interferometric observations, with only unresolved 21cm absorption detections \citep{MacHattie2014,Taft2019}. In this paper, we present observations of the galaxy in 21cm obtained with the MeerKAT telescope which provides increased surface brightness sensitivity as compared to interferometers previously available for \hi\ observations. In section \ref{sec:observations} we present the observations, in section \ref{sec:radio_cont} we characterize the radio continuum source in Haro 11, in section \ref{sec:HI-mass} and \ref{sec:HI-geom} we detail the neutral gas content and geometry of the system respectively, we discuss our findings in section \ref{sec:discuss}, and present our concluding remarks in section \ref{sec:conclusion}. 

Throughout this paper we assume a Hubble constant $\rm{H}_{0} = 67.4 \pm 0.5\, \rm{km}.\rm{s}^{-1}\rm{Mpc}^{-1}$  and a matter density $\Omega_{\rm{m}}= 0.315$ \citep{Planck_Collaboration2020}. Using the redshift of Haro 11 \citep[z=0.0206][]{Bergvall2000}, we derive a luminosity distance of 93 Mpc. Literature values mentioned here have been corrected for this value.

\section{Observations and data reduction} \label{sec:observations}

\subsection{MeerKAT 21cm observations} 
Haro 11 was observed with the MeerKAT interferometer as part of program SCI-20210212-AL-01 (P.I. Le Reste). The 8.8 hours of on-source integration time were split between two observing sessions (2021 Feb. 20 and 2021 Aug. 14). The correlator was configured in the C856M32k mode, wherein an 856 MHz wide bandpass is separated into 32,768 channels. J0408-6545 and J1939-6342 were used for primary flux and bandpass calibrations for the Feb 20 and the Aug 14 data, respectively; J0025-2602 was used as secondary calibrator in both observing sessions. To study the 21cm \hi\ line, a narrow (20 MHz) bandpass, centered on the recessional velocity of Haro 11 derived from optical emission lines \citep{Bergvall2000}, was extracted from the full dataset. The 26.123 kHz channel width produces a native velocity resolution of 5.5 km/s. Data reduction and calibration followed standard prescriptions in the CASA 5.6 environment \citep{CASA2022}. The calibrated dataset was continuum-subtracted using the task \texttt{uvcontsub}. Imaging was performed with CASA 6.4 using the \texttt{AUTO-MULTITHRESH} algorithm \citep{Kepley2020} within the \texttt{tclean} task. Three clean data cubes with different angular resolutions were produced by using tapering in the uv-plane, each with velocity resolution set to 10 km/s. The masks generated by \texttt{AUTO-MULTITHRESH} were visually examined, and deep cleaning (to 0.5 $\sigma$) was performed inside regions containing 21cm signal. Visual examination confirmed the presence of 21cm absorption in the cube, which had been detected by high angular resolution interferometric observations previously \citep{MacHattie2014}. The final datacubes have beam sizes and rms noise values as follows: 47.3\arcsec$\times$45.7\arcsec\ and $\sigma$ = 0.18 mJy/beam (``low resolution'', produced using a 2.5k$\lambda$ uv-taper); 24.6\arcsec$\times$21.7\arcsec\ and $\sigma$ = 0.16 mJy/beam (``medium resolution'', produced using a 7k uv-taper); 11.2\arcsec$\times$9.7\arcsec\ and $\sigma$ = 0.17 mJy/beam (``high resolution'', produced with no uv-taper).

The dedicated 21cm detection software Source FInding Algorithm \citep[\texttt{SoFIA}, version 1.3.3][]{Serra2015} was used to detect 21cm signal in the cubes. The Smooth+Clip algorithm was used with 4.5$\sigma$ threshold and median absolute deviation rms mode in order to include \hi\ absorption in the mask. Spatial smoothing with Gaussian kernels and spectral smoothing boxcar kernels (either no smoothing or $\sim30\,$km/s smoothing) were used. For each of the datasets, kernels with no smoothing, 10" smoothing, 23" smoothing and 46" smoothing were applied. Eye inspection of the masked cubes confirmed that the 21cm detection parameters enabled the recovery of all \hi\ emission and absorption in the cubes.

 \begin{figure*}[t]
    \centering
    \includegraphics[width=0.9\textwidth]{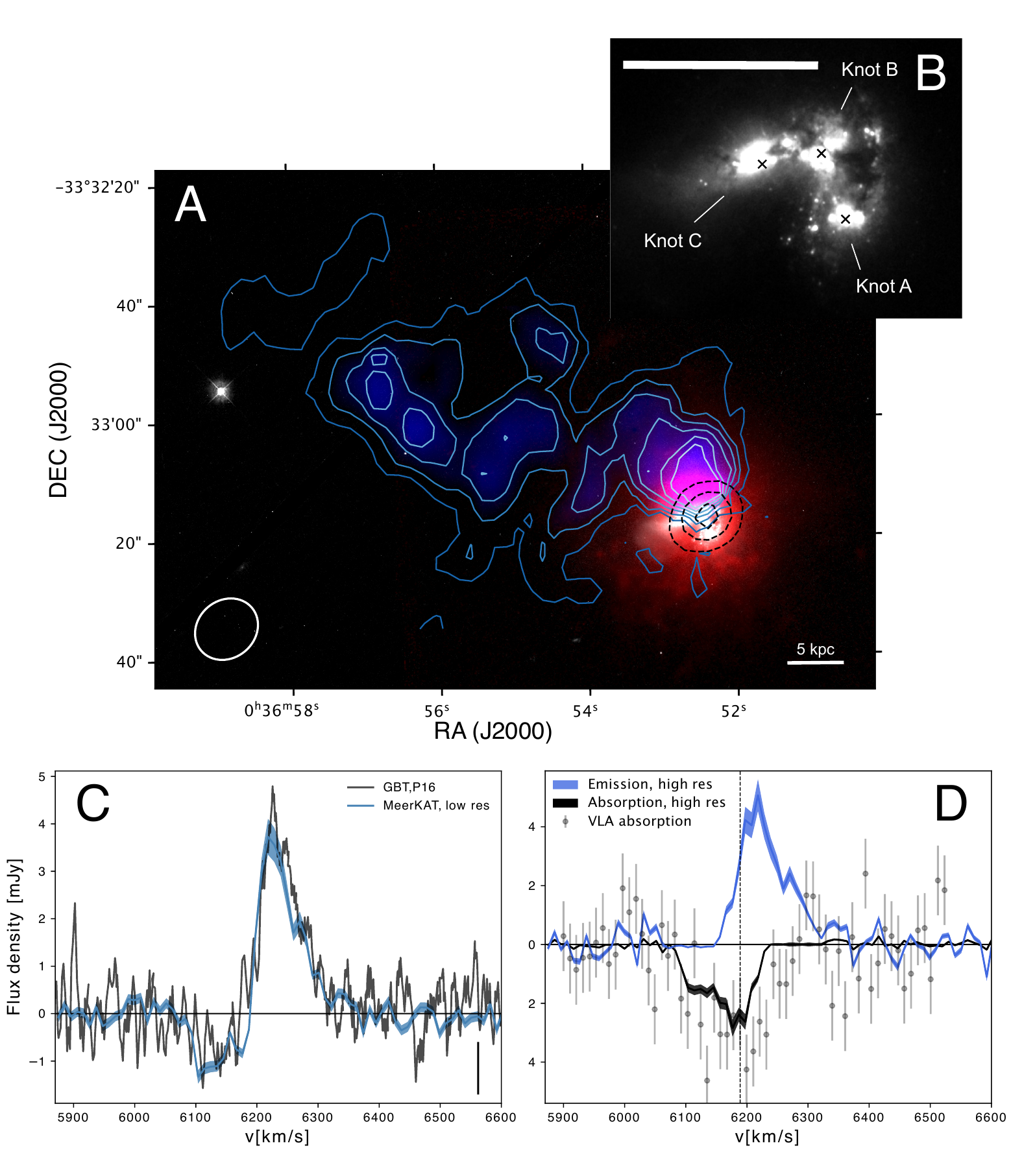}
    \caption{Neutral gas in Haro 11. Panel A: Colour-composite image of Haro 11 with 21cm contours. High resolution (10") MeerKAT 21cm emission
 is shown in blue with blue contours overlaid, MUSE H$\alpha$ emission is shown in red and HST optical stellar light observed using the F435W filter in white. The MeerKAT emission contours are shown with levels \nhi = \{0.7,1.4,2.1,2.8,3.4,4.1\}$\times10^{20}\,\rm{cm}^{-2}$ (0.007 to 0.04 Jy/beam.km/s), corresponding to the $5 \sigma$ to $30 \sigma$ levels, with lower $\sigma$ contours shown in darker shades of blue. MeerKAT 21cm absorption contours are overlaid in dashed black lines, with levels \{-0.3,-0.2,-0.1\} Jy/beam.km/s displayed. The MeerKAT synthesized beam is shown by a white ellipse in the lower left corner of the image. Panel B: Stellar light shown by the HST F435W image, with star-forming knots identified by crosses. In panel A and B, a white bar indicates the scale corresponding to 5 kpc. In all images, North is up and East is to the left. Panel C: Low angular resolution (47") MeerKAT 21cm spectrum integrated over all detected 21cm (emission and absorption), shown in blue. The GBT spectrum is shown in gray for comparison. A vertical black bar on the bottom right corner shows the typical error on the GBT spectrum. Panel D: MeerKAT 21cm high angular resolution (10") emission (blue) and absorption (black) spectra, integrated over the regions shown respectively by blue and black contours in panel A. The VLA absorption spectrum is shown with gray dots for comparison. The vertical dashed line shows the velocity centroid of ionized gas around the star-forming knots, where the sources emitting the absorbed radio continuum radiation are located.}
    \label{fig:21cm_global}
\end{figure*}

\subsection{VLA radio continuum data} 
In order to characterize the strong radio continuum source against which 21cm absorption occurs, we used archival radio continuum data. The data was taken by the National Radio Astronomy Observatory’s Karl G. Jansky Very Large Array (VLA) as part of project VLA/15B-197 (PI: Kepley). As part of the project, Haro 11 was observed with the S, X and Ka bands. The observing frequencies and parameters are given in Table \ref{tab:radio_cont_obs}. For all three bands observed, a flux/bandpass calibrator was observed once during each scheduling block and the phase calibrator was observed periodically. Additionally for the X and Ka-band data, a pointing scan was run at the beginning of a scheduling block. The data were calibrated using VLA pipeline version Pipeline-Cycle3-R1-B with CASA version 4.3.1 r32491 and imaged in CASA 6.2.1-7, using multi-term, multi-frequency synthesis \citep{Rau2011} with nterms=2 to account for the spectral curvature of the data and Briggs robust weighting of 0.5. In addition, the S-band data used w-projection to correct for the curvature of the sky over the larger field of view. We used self-calibration to improve the image dynamic range for all three bands. All three datasets had a phase self-cal performed. The Ka-band data had an additional scan-length amplitude self-cal. The final images were primary beam corrected accounting for the variation of the primary beam as a function of frequency across the band.
\begin{table*}
    \centering
    \begin{tabular}{l|ccc}
    \hline
        Band & S-band & X-band & Ka-band \\
        \hline
        Configuration & B & CnB & DnC \\
        Central Frequency [GHz] & 3 & 9.8 & 33.0 \\
        Bandwidth [GHz] & 1.75 & 4.0 & 7.9 \\
        Channel width [MHz] & 1 & 1 & 1  \\
        Number of channels &  128 & 128 & 128 \\
        Number of spectral windows & 16 & 32 &64  \\
        Flux/BP calibrator & 3C48 & 3C48 &3C48 \\
        Phase calibrator & J0024-4202-S & J0012-3954 &  J0012-3954\\
        Phase calibrator cadence [min] & 14 & 14  &  7\\
        Time on Source [hr]& 0.26 & 0.24 &  3.1\\
        Observing Dates & 2016-05-31 & 2016-05-05&  2016-01-10 to 2016-01-24\\
        Beam & 5.3" $\times$ 1.65" & 3.22" $\times$ 1.51"& 2.31" $\times$ 1.35" \\
        Image noise [mJy/beam] & 15.6 & 9.1& 6.7 \\
     \hline      
    \end{tabular}
    \caption{Very Large Array radio continuum observation parameters.}
    \label{tab:radio_cont_obs}
\end{table*}

\section{Radio continuum emission}
\label{sec:radio_cont}

\begin{figure*}
    \centering
    \includegraphics[width=\textwidth]{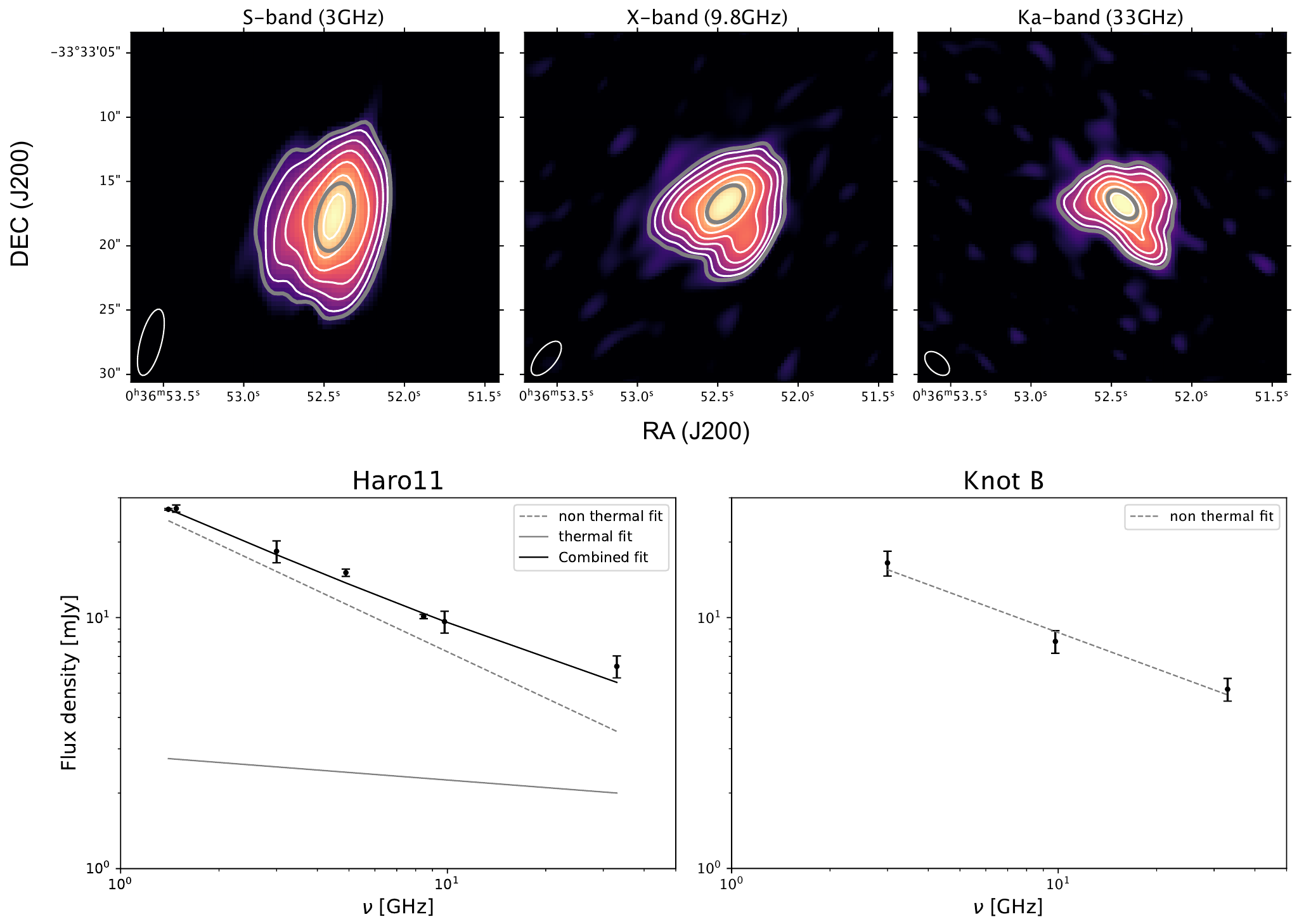}
    \caption{Radio continuum in Haro 11. The panels on the top row show, from left to right, the S-band (3 GHz), X-band (9.8 GHz) and Ka-band (33 GHz) continuum images. Intensities are displayed with a logarithmic stretch to highlight the full distribution of the radio continuum emission. The white contours displayed correspond to the {5,10,20,40,80,160} $\sigma$ levels. The synthesized beam is represented by a white ellipse in the bottom left corner. The contours used to extract the radio continuum flux from Haro 11 (3$\sigma$ level) and Knot B (Gaussian fit) are shown in gray. The bottom row panels show the SED for Haro 11 (left) and star-forming knot B only (right). For Haro 11, we fit the data using a combination of thermal and non-thermal radiation. The combined fit is shown in black, the thermal and non-thermal components are shown in gray solid and dashed lines respectively. For Knot B, we fit the data using the non-thermal emission only (gray dashed line), since we do not have enough data points to make a combined fit.}
    \label{fig:radio_cont}
\end{figure*}

We present the VLA radio continuum images of Haro 11 in the top panels of Fig. \ref{fig:radio_cont}. The radio continuum flux density of Haro 11 was calculated in the VLA archival images using the lowest contour shown on Fig.  \ref{fig:radio_cont}, corresponding to the 3$\sigma$ noise level. The flux density in all three bands observed with the VLA was supplemented by values taken from the literature \citep{MacHattie2014,Schmitt2006}. The flux density values for Haro 11 are presented in Table \ref{tab:radio_cont_Haro11}. 

While radio continuum emission is produced across the optical body of the galaxy, the strongest source is co-spatial with star-forming Knot B, and is unresolved, even with Ka-band imaging. In fact, the radio continuum emission in Haro 11 is completely dominated by the contribution from Knot B. We compare the radio continuum morphology and the location of the Knot B radio continuum emission with the optical image of Haro 11 in Appendix Fig. \ref{fig:Haro11_abs_centr}. X-Ray observations of the galaxy found two Ultra Luminous X-ray regions, respectively overlapping with Knot C and Knot B \citep{Gross2021,Prestwich2015}. These observations could indicate the presence of a low luminosity Active Galactic Nucleus (AGN) in the X-Ray region co-spatial with Knot B.  We extract the flux density in Knot B only in the VLA archival images. To do so, we fit a 2D Gaussian using the CASA task \texttt{imfit}, fixing the value of the peak and allowing for a zero-point offset. The flux density values of Knot B only are presented in Table \ref{tab:radio_cont_Haro11}. The error values are calculated using a conservative flux calibrator error of 10 \%, which drives the uncertainty. 

\begin{table*}
    \centering
    \begin{tabular}{c|ccccccc}
    \hline
        $\nu\,$[GHz] & 1.4 & 1.48 & 3 & 4.89 & 8.46 & 9.8 & 33 \\
        \hline
        S$_{\nu,\textrm{Haro11}}$ [mJy] & 27.0 $\pm$ 0.2 & 27.20.9$\pm$0.9 & 18.4$\pm$1.8 & 15.1$\pm$ 0.5 & 10.1 $\pm$ 0.2 & 9.6 $\pm$ 1.0 & 6.4 $\pm$ 0.6\\
         S$_{\nu,\textrm{KnotB}}$ [mJy]& - & - &16.5$\pm$1.7 & - & -& 8.0$\pm$0.8 & 5.1$\pm$0.5\\
        \hline
    \end{tabular}
    \caption{Radio continuum flux density values for Haro 11 and Knot B only.}
    \label{tab:radio_cont_Haro11}
\end{table*}
 We show the flux extraction regions used for the full galaxy and Knot B on Fig. \ref{fig:radio_cont}. We fit the radio Spectral Energy Distribution (SED) of Haro 11 using a combination of thermal and non-thermal radiation: 
$S(\nu) =  S_{nth} + S_{th} = c_1\, \nu^{-\alpha} + c_2 \nu^{-0.1}$. We use a least-square fitting approach weighted by errors and find the following parameters produce the best fit to the data: $c_1 = 29.9 \pm 4.5\, \rm{mJy}$, $\alpha =0.6 \pm 0.27$, $c_2 = 2.8 \pm 5.3\, \rm{mJy}$. Given the errors on the thermal emission coefficient, the radio continuum emission in Haro 11 is consistent mostly with non-thermal emission, and is dominated by the emission from Knot B. The non-thermal spectral slope is consistent with radio emission from star-forming regions \citep{Klein2018}. We fit the SED of Knot B with non-thermal radiation only, given the number of points is insufficient for a combined fit, and will lead to over-fitting. We find the following parameters minimize the chi-square: $c_1 = 26.3 \pm 5.1\, \rm{mJy}$, $\alpha = 0.5 \pm 0.1$. The flatter spectral index of Knot B is consistent within errors with that of the galaxy, and is consistent with that of star-forming regions. Therefore, we cannot conclude on the presence or absence of an AGN with the radio continuum observations.

\section{The neutral gas content of Haro 11} \label{sec:HI-mass}
The MeerKAT 21cm \hi\ image and spectra are presented in Fig. \ref{fig:21cm_global} with the H$\alpha$ emission of Haro 11 from MUSE \citep{Menacho2019} tracing ionized gas and optical stellar light from HST filter F435W imaging \citep{Ostlin2009}. We detect and resolve \hi\ in emission that was previously detected (Fig. \ref{fig:21cm_global}, panel C) by the Green Bank Telescope (GBT) \citep{Pardy2016} and detect the unresolved absorption component (Fig. \ref{fig:21cm_global}, panel D) detected by the VLA \citep{MacHattie2014,Taft2019}.

\subsection{\hi\ mass estimation} 
The MeerKAT 21cm high angular resolution cube was used for most of the \hi\ science applications, as the resolution allows to separate the 21cm emission and the significant 21cm absorption components. We have summarized the \hi\ properties of the galaxy in Table \ref{tab:hi_prop}, and detail the methods used to obtain these values below. To estimate the gas mass associated with 21cm emission only, the absorption component was manually masked from the cube. A double Gaussian function was fitted to the emission profile using least-square fitting weighted by errors. The fitted profile was integrated, yielding the 21cm emission flux $\rm{S}_{HI,em}= 0.391 \pm 0.042\, \rm{Jy.km/s}$. 
Flux error estimation accounts for a conservative flux calibrator error of 10\%, the variance in the cube and the uncertainty on fitting parameters. To determine the emitting gas mass, we assume that the gas is optically thin and use the standard equation \citep[e.g.][]{Roberts1962}
\begin{equation}
\rm{M}_{HI} [M_{\odot}]= 2.36 \times 10^5 \rm{D}^2\rm{[Mpc]}\, \rm{S}_{HI}\rm{[Jy.km/s]}
\end{equation}

This yields a new mass estimate for the \hi\ in emission 
$\rm{M}_{HI,em} = 7.99 \pm 0.85 \times 10^8\, \rm{M}_{\odot}$, a value larger than that of $5.7\pm0.8 \times 10^{8}\, \rm{M}_{\odot}$
previously calculated with the single dish measurement \citep{Pardy2016}. Values differ owing to the possibility to differentiate between the emission and absorption components with the MeerKAT observation, which average each other out in the unresolved GBT observation.

The gas mass associated with 21cm absorption was estimated separately, by inverting the absorption component mask to mask all emission. The calculation requires an assumption on the volume covered by the gas, and is a function of the area covered by the gas $\rm{A}_{\rm{gas}}$, the column density $\rm{N}_{HI}$, the mass of the hydrogen atom $\rm{m}_{\rm{H}}$ and the volume filling factor, which is a function of the gas covering fraction f. We assume the same volume filling factor $\rm{f}^{3/2}$ as in the previous 21cm absorption study of Haro 11 by \citet{MacHattie2014}, such that 
\begin{equation}
\rm{M}_{HI,abs} = \rm{N}_{HI,abs}\,A_{gas}\,m_{H}\,f^{3/2}
\label{eq:MHI_abs}
\end{equation}
The absorbing \hi\ gas column density $ \rm{N}_{HI,abs}$ can be written as a function of spin temperature $\rm{T}_{\rm{s}}$ and the optical depth $\tau(v)$ as:
\begin{equation}
\rm{N}_{HI,abs}\,\rm{[cm^{-2}]} = 1.823 \times 10^{18}\, \rm{T}_{\rm{s}}\,\rm{[K]} \int_{v} \tau(v) \rm{d}v\, \rm{[km/s]}
\end{equation}
The optical depth can be expressed as a function of the observed change in flux density due to the absorption $\Delta$S, the continuum flux density $\rm{S}_c$ and the covering fraction of the absorbing gas f as
\begin{equation}
\tau(v) = - \rm{ln}(1+\Delta S(v)/(f \,\rm{S}_c)) 
\end{equation}
    The absorbing gas mass is subject to significant uncertainties, due to the unknown size of the absorbing component, spin temperature and covering fraction. The radio continuum images (see Fig. \ref{fig:radio_cont}) indicate that radio emission is produced across the optical body of the galaxy. While the radio continuum is dominated by the contribution of Knot B, the distribution of gas covering the knots is unknown, since neither our observation nor the previous VLA absorption study \citep{MacHattie2014} resolve the absorption. For this reason, assuming that all of the gas is distributed in front of Knot B would likely lead to a severe underestimation of the \hi\ mass of the absorbing gas. We present the calculation for this hypothesis in Appendix section \ref{sec:mass_abs_kb}, but follow a more conservative approach that is informed by the observations at hand in the following. We use the beam size of the VLA observations (7.27" $\times$10.07") \citep{MacHattie2014} as an upper limit to the area covered by the absorbing gas. Observations and modeling of the interstellar medium in Haro 11 indicate that the spin temperature is within the range $\rm{T}_{\rm{s}}= 91-200 \,\rm{K}$ \citep{MacHattie2014,Cormier2012}. To constrain the covering fraction, we use the values derived from the measurement of the $\rm{Si}_{\textsc{ii}}$ UV absorption line, which is often used as a proxy for studying the neutral gas in galaxies. We use the average covering fraction value of all three knots at deepest absorption velocity $v_{min}$, yielding $ \rm{f} = 0.82$ \citep{Ostlin2021}. Using these parameters, we find an absorbing gas mass $\rm{M}_{HI,abs} = 3.10 \pm 1.16 \times 10^{8} \rm{M}_{\odot}$.

The total \hi\ mass of Haro 11 is calculated by adding the 21cm emitting gas mass to the the absorbing gas mass, with errors taking into account the uncertainties on both values, dominated by the 21cm absorbing gas mass uncertainty. This yields a total neutral hydrogen gas mass $\rm{M}_{HI} = 1.11 \pm 0.20 \times 10^{9}\, \rm{M}_{\odot}$.
This value is higher than previous estimates. Furthermore, the 21cm image shows the presence of gas with column density well above $\rm{N}_{\rm{HI}}=2\times10^{20}\,\rm{cm}^{-2}$ (see Fig. \ref{fig:21cm_global}), the empirical efficient self-shielding limit of \hi\ \citep{Kanekar2011}. Nevertheless, the \hi\ mass remains smaller than the ionized gas mass \citep{Menacho2019}  
$1.8  \times 10^{9}\, \rm{M}_{\odot}$
and the stellar mass \citep{Ostlin2001} of the galaxy
$\rm{M}_{*} = 1.6^{+2.1}_{-0.6}  \times 10^{10}\, \rm{M}_{\odot}$, meaning that neutral gas has been efficiently converted into stars and ionized, likely helping the escape of LyC photons. Most importantly, the neutral gas is strongly offset from LyC production regions, which we discuss hereafter.

\begin{table}
    \centering
    \begin{tabular}{c|c}
    \hline
       $\rm{S}_{HI,em}$ &$0.391 \pm 0.042\, \rm{Jy.km/s}$ \\
       $\rm{M}_{HI,em}$ & $7.99 \pm 0.85 \times 10^8 \rm{M}_{\odot}$ \\
       $\rm{M}_{HI,abs}$ & $  3.10 \pm 1.16 \times 10^{8} \rm{M}_{\odot}$ \\
        $\rm{M}_{HI,tot}$ & $1.11 \pm 0.20 \times 10^{9} \rm{M}_{\odot}$ \\
        $\rm{M}_{HI,tail}$ & $6.72 \pm 0.88\times 10^8 \rm{M}_{\odot}$ \\
    \hline
    \end{tabular}
    \caption{Neutral gas properties of Haro 11 and different components in the galaxy.}
    \label{tab:hi_prop}
\end{table}

\section{Neutral gas geometry}  \label{sec:HI-geom}

\subsection{Global \hi\ distribution} 

The \hi\ emission is offset from the main body of the galaxy, in an elongated structure that is $\sim40\,\rm{kpc}$ long when projected on the plane of the sky (Fig. \ref{fig:21cm_global} panel A). The 21cm emission components in individual velocity channels are found to be connected spatially and spectrally (especially visible in Appendix Fig \ref{fig:chanmap_interm_res}, but also \ref{fig:chanmap_high_res}). Additionally, the \oi/\ha\ line ratio map, where elevated values trace shocks, shows the largest values (\oi/\ha$>0.1$) to the North East of the galaxy \citep{Menacho2019}, towards the direction of the extended neutral gas structure. We have shown the \ha\ and \oi/\ha\ maps of Haro 11 with \hi\ contours overlaid in Appendix Fig. \ref{fig:MUSE_maps}. These large values indicate that a shock is occurring at this location. This supports a view with two kinematically distinct gas components, instead of the neutral gas structure being e.g. the result from a continuous outflow from the center of the galaxy. Together, these observations are consistent with the \hi\ emission structure being a tidal tail from a merger. 

The merger nature of Haro 11 had already been demonstrated by ionized gas kinematics studies \citep{Ostlin2001,Ostlin2015}, but the impact of the interaction on the neutral gas distribution was previously unknown. In fact, the neutral gas structure in Haro 11 strongly resembles a smaller version of the Southern tidal tail in the more massive Antennae galaxy merger \citep{Hibbard2001}, to which Haro 11 has been compared previously due to striking similarity in ionized gas kinematics \citep{Ostlin2015}. To further analyze how the merger impacts the neutral gas distribution and star formation in Haro 11 as a function of time, our team has developed simulations of the merger that will be presented in a forthcoming paper \citep[][in prep.]{Ejdetjarn}. These simulations use the RAMSES code \citep{Teyssier2002} and reproduce the \hi\ structure presented here, as well as observations of the stellar populations in the Knots \citep{Sirressi2022}. The Haro 11 system was reproduced by the merger of two disc galaxies with respective stellar mass $M_{*,1} = 8\times 10^{9} M_{\odot}$ and $M_{*,2} = 4.6\times 10^{9} M_{\odot}$. The single tidal tail is reproduced in a scenario where the galaxies rotate in opposite ways, with the most massive galaxies having prograde motion, while the least massive one has retrograde motion.  According to the simulation, the tail was formed in a period of over 250 Myr after the first close-by interaction between the galaxy nuclei. Haro 11 is currently observed 10 to 20 Myr after the second close interaction and about 20 Myr before nuclear coalescence. In the simulation, Knot B and C correspond to the nuclei of the initial galaxies, while Knot A is formed in the merging process through tidal motions of the stars.

The asymmetric neutral gas distribution resulting from the merger in Haro 11 likely facilitates the escape of LyC emission in a large fraction of the solid angle, including our own line of sight, and prevents it from other parts. If the tidal tail had been oriented towards the Earth, LyC emission from Haro 11 would most likely not have been detected. The gas with projected  column density corresponding to the empirical efficient self-shielding limit of hydrogen $2\times10^{20}\text{cm}^{-2}$ \citep{Kanekar2011} (corresponding approximately to the third contour on Fig. \ref{fig:21cm_global}) covers an angle of 135\textdegree\ from the center of the galaxy. Making the simplest assumption to estimate the solid angle fraction covered by \hi\ with sufficient column density to efficiently self-shield, which is radial symmetry of the \hi\ gas along the tidal tail axis, we find:
\begin{equation}
\Omega_{\text{\hi}} = \int_{\theta_0}^{\theta_f}\sin\theta\textrm{d}\theta\int_{\phi_0}^{\phi_f}\textrm{d}\phi = \int_{0}^{\frac{3\pi}{4}}\sin\theta\textrm{d}\theta\int_{\frac{\pi}{4}}^{\pi}\textrm{d}\phi \simeq 0.32\,\Omega_{\text{tot}}
\end{equation}
On large scales characterizing the circumgalactic medium, the merger has thus cleared almost 70\% of the solid angle from neutral gas with sizeable column density in Haro 11. Mergers are dynamical phenomena operating on timescales on the order of hundreds of million years, therefore the position and solid angle covered by neutral gas will change as a function of time. As a result, ionizing radiation escaping from a merger is necessarily strongly anisotropic and varies with time. In the absence of information on the 3D distribution of the gas, the estimation made with present observations is flawed. More precise calculations of the fraction of solid angle covered by neutral gas from each knots, and its evolution as a function of time will be presented in the paper describing simulations of the merger in Haro 11 \citep[][in prep.]{Ejdetjarn}.

\begin{figure}
    \centering
    \includegraphics[width=0.49\textwidth]{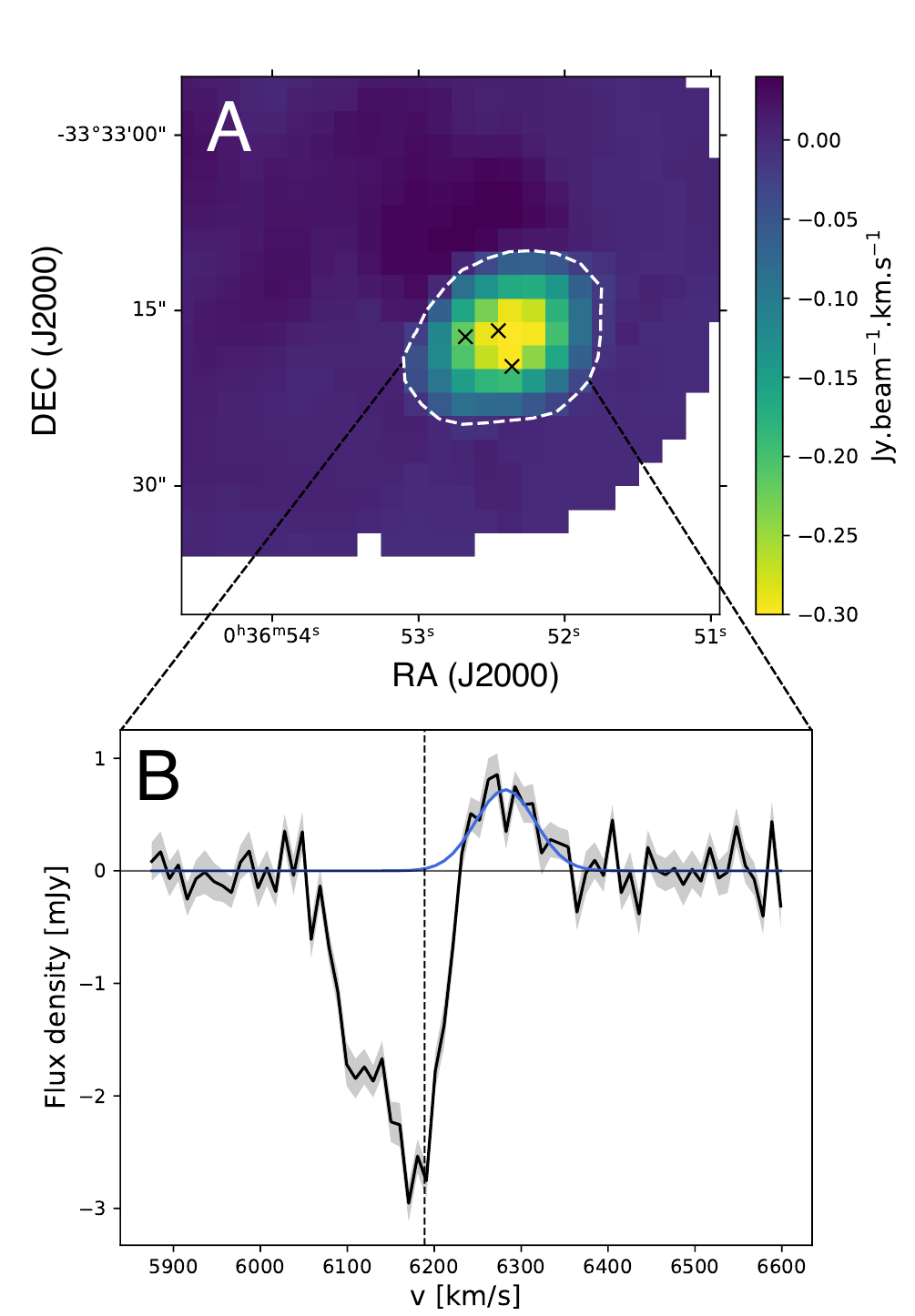}
    \caption{Neutral gas around the LyC production sites. Panel A: MeerKAT High angular resolution 21cm integrated flux density map. The dashed white line corresponds to the aperture used to extract the spectrum shown in the lower panel, which traces the limits of the absorption component, determined by collapsing the absorbing gas mask in velocity and applying it as an aperture on the non-masked cube. The black crosses indicate the position of the star-forming knots. Panel B: 21cm spectrum of Haro 11 at the location of the absorption component, shown in black. The blue line shows the Gaussian fit to the emission component seen at this location. The dashed black line shows the velocity centroid of the H$\alpha$ line extracted in the same aperture, assumed to indicate the central velocity of the ionized gas around the star-forming regions.}
    \label{fig:21cm_abs_em}
\end{figure}

\subsection{\hi\ geometry around LyC-emitting regions} 

To assess the content and kinematic structure of the neutral gas around the LyC production regions, we extract a spectrum from the non-masked cube at the location of the star-forming knots. This extraction region corresponds to the location of the 21cm absorbing gas component, since it is unresolved and covers the optical body of Haro 11. The spectrum and extraction aperture are shown on Fig. \ref{fig:21cm_abs_em}. 
We find that some 21cm emission is co-spatial with the star-forming regions and absorption feature, however the channel map in Appendix Fig. \ref{fig:chanmap_high_res} indicates the components are separated in velocity. We have shown the velocity centroid of the ionized gas emission, which we assume is similar to the velocity of the source emitting the radio continuum emission (consistent with radio emission from star-forming regions). We find that the peak of the 21cm emission component is redshifted by $92\pm5\,\rm{km}.\rm{s}^{-1}$. 

Precisely determining the position of \hi\ gas around the optical body of galaxies is generally difficult. Here however, we can use reasoning guided by the nature of the radio continuum emission and the velocity distribution of the absorption and emission components to gain some insight on the \hi\ distribution. Analysis of the radio continuum emission indicates the radio continuum is consistent with that of star-forming regions (see section \ref{sec:radio_cont}). \hi\ gas that is absorbing radio continuum at 21cm is necessarily in front of the radio continuum source, which we assume corresponds to star forming regions in Haro 11. Thus, the blueshift of the absorbing gas compared to the ionized gas centroid indicates that the absorbing gas is outflowing. The maximum velocity of the outflow as compared to the velocity centroid of the H$\alpha$ line is 97 km/s. Multiphase outflows were already identified around Knot B and Knot A through analysis of the kinematics of emission lines tracing the ionized gas and UV absorption lines from metals tracing a mix of neutral and ionized gas \citep{Sirressi2022}. This study found a maximum velocity of $\sim$400 km/s, much larger than the value found with 21cm absorption. However, metal UV absorption lines do not trace a perfectly neutral gas phase, and the excellent agreement between the kinematics from emission lines tracing ionized gas and UV absorption lines in this study could indicate that a significant fraction of the metals are found in the ionized gas phase. Regardless, the blueshift of 21cm absorption confirms that a neutral gas outflow is occurring in Haro 11.

Having identified that the absorbing gas is outflowing, we then consider the structure of the gas emitting in 21cm that overlaps with the position of Haro 11. This 21cm emitting gas is redshifted. While the emitting gas could possibly be located in front of the optical body of the galaxy and inflowing, it is less physically plausible than a scenario where it is behind the star-forming regions. We can reasonably expect the outflow impacting the neutral gas in front of the star-forming regions to impact gas behind the optical body of the galaxy as well. The redshift of the 21cm-emitting component compared to the centroid of ionized gas is consistent with the picture where this component is outflowing on the back-side of the galaxy, while the absorbing component is outflowing in front of it on our line of sight. 
\begin{figure}
    \centering
    \includegraphics[width=0.45\textwidth]{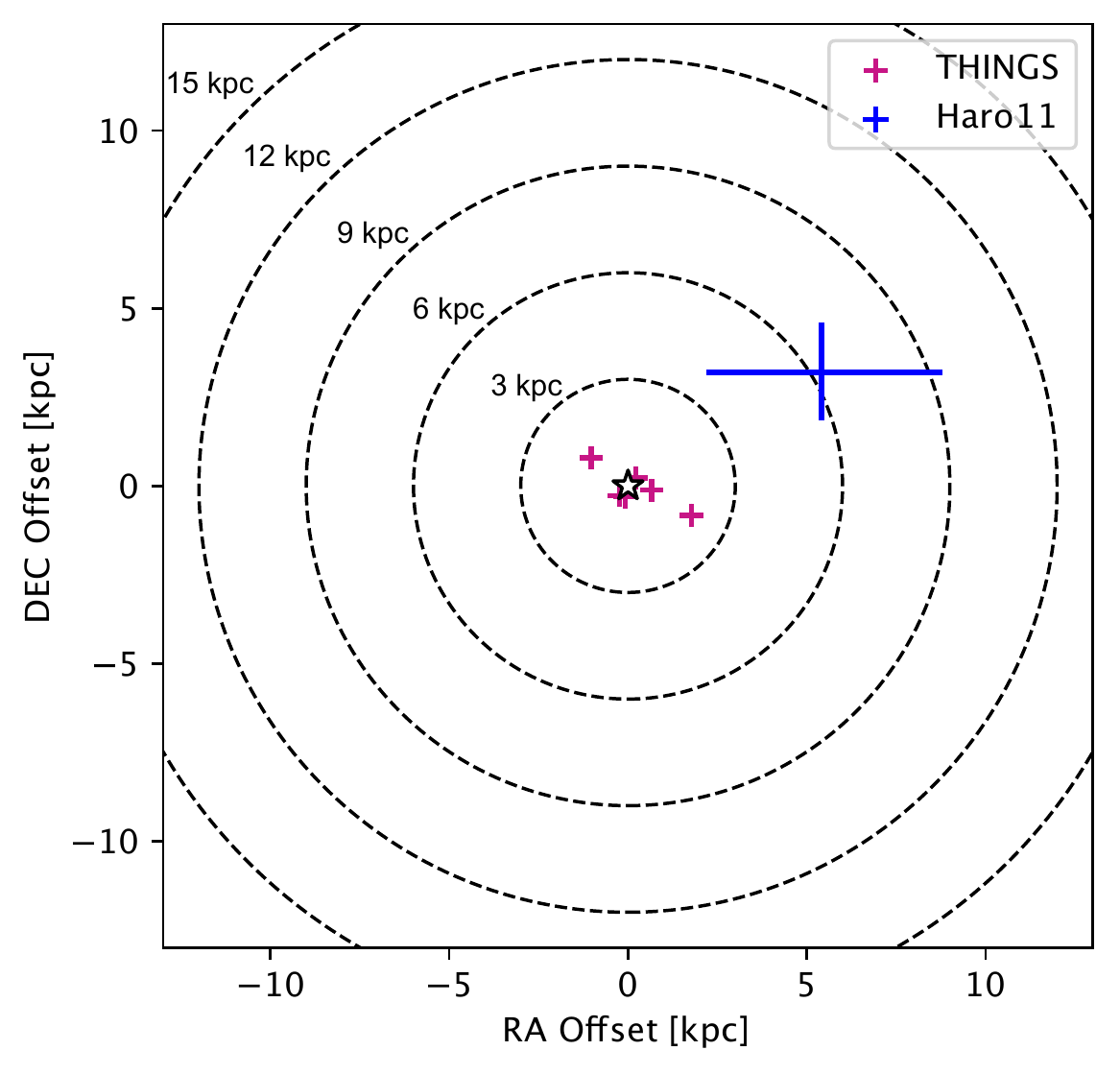}
    \caption{Relative 21cm centroid offset compared to the galaxy optical coordinates. The center of the galaxies is shown by the white star in the center. The position of the 21cm emission centroid is shown by purple crosses for galaxies in the THINGS sample \citep{Walter2008} with stellar masses in the range [$10^{10}-10^{11}\,\rm{M}_{\odot}$]. The position of the Haro11 \hi centroid is indicated by the blue cross, with error bars taking into account the uncertainty on the 21cm absorption component contribution to the \hi\ mass. Dashed lines indicate offsets from the center in multiples of 3 kpc.}
    \label{fig:21cm_offset}
\end{figure}

 We fit the 21cm emission component co-spatial with the absorption feature and optical body of the galaxy with a Gaussian using least-square fitting weighted by errors and integrate the Gaussian profile. We find a flux of $0.062\pm0.011\, \rm{Jy\,km/s}$ ($1.27 \pm 0.23 \times 10^8\, \rm{M}_{\odot}$), which corresponds to 16\% of the total 21cm \textit{emission} component flux of the galaxy. Most of the \hi\ gas that \textit{emits} in 21cm is thus found in the tidal tail, away from the star-forming knots. When taking into account the absorbing gas mass which is in front of the galaxy, we find that 25 to 64\% of the total neutral gas mass in the galaxy is offset from the ionizing emission production regions on our line of sight, mostly due to merger-driven interaction. We note that the \hi\ mass seen in emission overlapping with the body of the galaxy (outside of the tidal tail) could be underestimated by our observation, which we discuss in section \ref{sec:caveats}. Nevertheless, the tail contains a significant \hi\ mass of $6.72 \pm 0.88\times 10^8 \rm{M}_{\odot}$. In the following, we compare the \hi\ offset in Haro 11 to typical values from normal star-forming galaxies.

\subsection{\hi\ offset calculation} 
We observe an offset of the 21cm emitting gas from the center of the galaxy. In order to estimate if this offset is significant, we calculate the centroid coordinates $x_c,y_c$ of the 21cm emission and 21cm absorption separately, using the following expression, where $x_i$ and $y_i$ are the pixel coordinates on the $x$ and $y$ axes and $I_i$ the pixel value at $x_i,y_i$:

\begin{equation}
x_c = \frac{ \sum_i x_i\times I_i }{ \sum_i I_i } \ ,\ y_c = \frac{\sum_i y_i\times I_i}{\sum_i I_i}
\label{eq:centroid}
\end{equation} 
The centroid of 21cm mass in Haro 11 is then calculated by taking the average of the 21cm emission and absorption centroid coordinates, weighted by the mass fraction in each component, and taking into account the uncertainties on both mass values. 

We investigate how this neutral gas offset compares with galaxies of similar stellar mass in the THINGS sample of nearby galaxies \citep{Walter2008,DeBlok2008}. Specifically, we select galaxies with stellar masses $10^{10}\,\rm{M}_{\odot}<M_*<10^{11}\,\rm{M}_{\odot}$ in \citet{DeBlok2008}, and calculate the 21cm emission centroid in the publicly available moment-0 maps with robust weighting using equation \ref{eq:centroid}.  We derive the 21cm emission centroid of NGC 925, NGC 2903, NGC 3198, NGC 3621, NGC 4736, NGC 6946\footnote{NGC 3031 is also in the stellar mass range considered, but imaging artefacts \citep{Walter2008} prevent an accurate calculation of the 21cm emission centroid.}, and compare it to that of Haro 11, accounting for the large uncertainty on the absorbing gas mass (see Table \ref{tab:21cm_centroids}).

We find an average offset of 0.82 kpc for the THINGS galaxies, with all galaxies having 21cm emission centroids within 2 kpc of the optical coordinates. In contrast, the 21cm centroid in Haro 11 is offset by 6.28$^{+3.64}_{-3.41}$ kpc, 4 to 12 times larger than the average centroid offset for THINGS galaxies. We have also quantified the offset to individual Knots: the centroid of the gas is offset by 8.89$^{+3.64}_{-3.41}\,$kpc,
7.71$^{+3.64}_{-3.41}\,$kpc, and 6.46$^{+3.64}_{-3.41}\,$ kpc respectively from Knot A, B and C.
We have shown the relative offset of the 21cm centroid compared to the central galaxy coordinates in physical units on Fig. \ref{fig:21cm_offset}. The 21cm centroids are shown overlaid on 21cm emission moment-0 maps for each galaxy in Appendix Fig. \ref{fig:21cm_centroid_images}. In general, neutral gas does not perfectly overlap with stellar regions in galaxies. However, the neutral gas in Haro 11 has undergone a bulk offset from the center of the galaxy, where the LyC-emitting regions are located. This global offset caused by a merger interaction is likely facilitating the escape of LyC photons from the center of the galaxy into our line of sight.

\begin{table}
    \centering
    \begin{tabular}{l|c}
    \hline
    ID & 21cm centroid offset [kpc]\\
    \hline
       Haro 11  & $6.28^{+3.64}_{-3.41}$ \\
       Haro 11, em  & 16.51 \\
       Haro 11, abs  & 2.04 \\
        NGC 925 & 1.30\\
        NGC 2903 & 0.32\\ 
        NGC 3198 & 0.31\\ 
        NGC 3621 & 0.37\\  
        NGC 4736 & 0.67\\  
        NGC 6946 & 1.96\\  
       \hline
    \end{tabular}
    \caption{Distance offset of the 21cm emission centroid to the optical coordinates of the galaxy. The uncertainties on the centroid offset in Haro 11 are due to the uncertainties of the absorbing gas mass located in front of the galaxy.}
    \label{tab:21cm_centroids}
\end{table}

\subsection{Caveats} 
\label{sec:caveats}
While the MeerKAT observation improves upon previous work on the neutral gas mass and distribution of Haro 11, we note that the 21cm absorption remains unresolved. Similarly to the way in which the emission and absorption cancelled each other out in the GBT observation, causing the absorption feature to be missed by previous single dish study, the MeerKAT observation could be missing part of the mass around the center of the galaxy due to angular resolution limits. In order to improve upon the \hi\ mass measurement obtained here, observations with both excellent sensitivity and resolution close to the angular size of the knots ($\sim2"$) will be required. Observing a deeper absorption feature at lower angular resolution would likely not affect the \hi\ mass in absorption significantly due to the decrease in mass when considering smaller gas-covering areas (see Equation \ref{eq:MHI_abs} and Appendix section \ref{sec:mass_abs_kb}). However, this could cause an increase of the \hi\ mass in the neutral gas component seen in emission that is overlapping with the optical body of the galaxy. Nevertheless, the offset of a large amount of \hi\ gas from our line of sight certainly facilitates the escape of LyC photons in our direction. If this mass was instead distributed uniformly within the area of the beam, the additional column density along the line of sight would be $1.8\times10^{21}\textrm{cm}^{-2}$. If the mass in the tail was distributed in a larger area corresponding twice the beam size, as one could reasonably expect given neutral gas usually extends beyond the optical body of normal galaxies, the corresponding column density would be $4.5\times10^{20}\textrm{cm}^{-2}$. Even when assuming that the LyC photons would "see" only half of this column density, these values are optically thick to LyC radiation by several orders of magnitude. Therefore, the tidal displacement of the gas outside of our line of sight effectively lowers the column density seen by LyC photons, and facilitates the ionizing radiation escape into the intergalactic medium.

\section{Discussion} \label{sec:discuss}

The \hi\ morphology and kinematics of Haro 11 demonstrate that large-scale pathways devoid of neutral gas with significant column density exist around this galaxy, likely facilitating LyC escape from the circumgalactic medium on several lines of sight.
 These observations provide a potential link between dwarf galaxy mergers and the detection of LyC emission from galaxies. Galaxy mergers in low mass systems could play several roles in facilitating the escape of LyC emission from galaxies. First, mergers create multiple star formation bursts \citep{Lahen2020} during the timescale of the interaction by repeatedly compressing the gas at the center of the galaxy. These bursts create numerous massive stars, which produce the bulk of LyC emission in galaxies; thus, mergers increase the intrinsic LyC photon production. Second, starbursts are also responsible for the intense feedback that creates ionized channels enabling the escape of LyC photons from their immediate environment \citep{Borthakur2014,Clarke2002}. Finally, since merger interactions lead to the efficient conversion of gas into stars in the central parts of galaxies, and the ionization of remaining \hi\, or displacement of it in tidal structures away from the star-forming regions \citep{Georgakakis2000}, they effectively lead to a bulk offset of the material inhibiting LyC escape from the center of interacting galaxies. This effect, imaged here in a galaxy that is close to nuclear coalescence, is likely to strongly depend on the timescale of the merger interaction and the viewing angle of the merging system, with a fraction of the lines of sight being obscured by tidal structures. However, by creating regions depleted of \hi\ on large scales, galaxy mergers are interesting mechanisms that can facilitate the anisotropic escape of LyC photons out of the interstellar medium and into the intergalactic medium.
 Other environmental effects, such as ram pressure stripping of neutral gas in high density environments, produce similar effects on the neutral gas of galaxies. These processes can also promote star formation on the edges of stripped tails, which could lead to LyC escape \citep[e.g.][]{Kenney2014}. However, there has been no observation of LyC emission from galaxies undergoing such environmental effect so far.
 
In the local Universe, two other LyC-emitting galaxies have been detected that are close enough to be imaged with interferometers: Tol 1247-232 and Mrk 54 \citep{Leitherer2016}. Both of these galaxies show prominent merger morphologies in the optical, however their neutral gas content is significantly different, with Mrk 54 having a high \hi\ mass $\rm{M}_{\rm{HI}} = 1.6 \pm 0.2 \times 10^{10}\, \rm{M}_{\odot}$ \citep{Haynes2018}, while Tol 1247-232 has not yet been detected in 21cm, and has currently only an upper limit on the mass $\rm{M}_{\rm{HI}} < 1 \times 10^{9}\, \rm{M}_{\odot} $ \citep{Puschnig2017}.
Given the morphology of these galaxies in the optical, it is likely that a fraction of their \hi\ gas has been removed by tidal interactions from the lines of sight where LyC is emitted. This would facilitate the ionization by the starburst and explain how LyC emission can escape from environments with such different neutral gas properties.

While they are not directly detected, many galaxies are considered to be LyC candidates due to their peculiar \lya\ line profile shapes or their high [OIII]/[OII] ratios, which indirectly trace LyC escape \citep{Izotov2018,Flury2022a,Verhamme2015,Flury2022b}. Among these candidates, many display signs of ongoing merger events. Green pea galaxies are a class of objects considered excellent analogs of high redshift LyC-emitting galaxies \citep{Izotov2016,Cardamone2009}. Single dish 21cm line measurements of green pea galaxies have suggested that galaxies with high [OIII]/[OII] ratios are less likely to be detected in 21cm, potentially indicating that galaxies with low \hi\ mass are more likely to leak LyC radiation \citep{Kanekar2021}. However, about a fifth of the sample of green pea galaxies studied in 21cm have neutral gas and galaxy properties indicative of either recent gas accretion or the presence of a gas-rich companion. Recently, \hi\ imaging of the green pea galaxy J0213+0056 has shown that a merger could explain \lya\ escape in the galaxy, and could potentially lead to LyC leaking \citep{Purkayastha2022}. One study of the environment of a small sample of green pea galaxies using optical emission lines has found that these galaxies do not have a higher companion fraction than less star-forming line-and continuum-selected galaxies at comparable stellar mass \citep{Laufman2022}. However, this result does not exclude these objects being the product of mergers close to nuclear coalescence, which could be investigated by studies of low surface brightness tidal features in these galaxies.

The galaxy merger rate is difficult to measure in the early Universe, however a few observational studies have found indications of an increase in merger rate at redshift 4 and up to 6 \citep{Tasca2014,Duncan2019}. From the simulation perspective, galaxy mergers also seem to be a promising process facilitating the escape of ionizing LyC radiation during the epoch of reionization. Indeed, cosmological hydrodynamical simulations predict an increase of mergers with increasing redshift \citep{Rodriguez-Gomez2015}. Semi-analytic galaxy formation models also show that galaxies found in dense environments reside in larger ionized regions \citep{Qin2022}, with galaxies having neighbors being more likely to show \lya\ emission at z=8. However, the environment of galaxies leaking LyC emission has not yet been studied in a systematic way, leaving uncertainties on their possible contribution to Reionization. 

\section{Conclusion} \label{sec:conclusion}
We have observed the neutral gas reservoir of the LyC-emitting galaxy Haro 11 with the 21cm \hi\ line. We detect a previously known unresolved 21cm absorption component overlapping with the galaxy, and resolve the 21cm emission, hereby providing the first map of the material preventing ionizing radiation escape in a LyC emitter. By calculating the mass in the 21cm absorbing and emitting components separately, we find a neutral gas mass $\rm{M}_{HI} = 1.11 \pm 0.20 \times 10^{9}\, \rm{M}_{\odot}$, larger than previous estimates from unresolved measurements in which these components averaged each other. We also find that the neutral gas is distributed in a $\sim40\,\rm{kpc}$-long tidal tail resulting from a merger interaction. This gas tail connects with the optical body of the galaxy, around which the neutral gas is outflowing. We find that $44^{+20}_{-19}$\% of the neutral gas mass resides in the tidal tail, and the gas has undergone a bulk offset of 
$6.28^{+3.64}_{-3.41}\, \rm{kpc}$ from the center of the galaxy, about 7 times larger than the mean offset between optical coordinates and 21cm emission for galaxies of similar mass in the THINGS survey. Additionally, we estimate that on large scale, the merger interaction has cleared about 70\% of the solid angle from dense gas with the ability to self-shield against ionizing radiation ($\rm{N}_{HI} > 2\times10^{20}\, \rm{cm}^{-2}$) by displacing a large fraction of the gas into the tidal tail and ionizing remaining gas surrounding the center.

The merger interaction is likely the mechanism which allows us to observe ionizing radiation from Haro 11, since it has increased the star formation rate in the center of the galaxy, leading to increased LyC production and to the creation of ionized channels, and has offset a large fraction of the material which inhibits LyC escape from star-forming regions from our line of sight.  
While mergers are likely not the only mechanism responsible for LyC escape from galaxies, they are an effective process that unites several of the conditions required for ionizing photons to escape to the intergalactic medium. Their contribution to reionization has not yet been evaluated in detail. The impact of galaxy environment on LyC escape should be assessed from the point of view of simulations and observations alike.

\section*{Acknowledgements}

ALR thanks Christian Binggeli, Alba Covelo Paz and Mohammad Javad Shahhoseini for their contributions to the pilot 21cm interferometric observing proposal for Haro 11.
JMC and JLI are supported by NSF/AST-2009894. JMC and SHT acknowledge support from Macalester College.
MJH is fellow of the Knut \& Alice Wallenberg foundation.
AA acknowledges financial support from the Swedish Research Council (VR) under grant 2021-05559.
G\"{O} acknowledges financial support from the Swedish Research Council (VR) and the Swedish National Space Agency (SNSA).
The MeerKAT telescope is operated by the South African Radio Astronomy Observatory, which is a facility of the National Research Foundation, an agency of the Department of Science and Innovation.
The National Radio Astronomy Observatory is a facility of the National Science Foundation operated under cooperative agreement by Associated Universities Inc.
This study uses observations made with ESO Telescopes at the La Silla Paranal Observatory under programme IDs 094.B-0944(A) and 096.B-0923(A).
This work made use of THINGS, 'The HI Nearby Galaxy Survey' \citep{Walter2008}.

\section*{Data Availability}

The datasets generated and analysed in the current study will be made publicly available in the SARAO DataCite repository.


\bibliographystyle{mnras}
\bibliography{bibliography}



\appendix

\section{Optical and radio continuum emission comparison}
\begin{figure}
    \centering
    \includegraphics[width=0.49\textwidth]{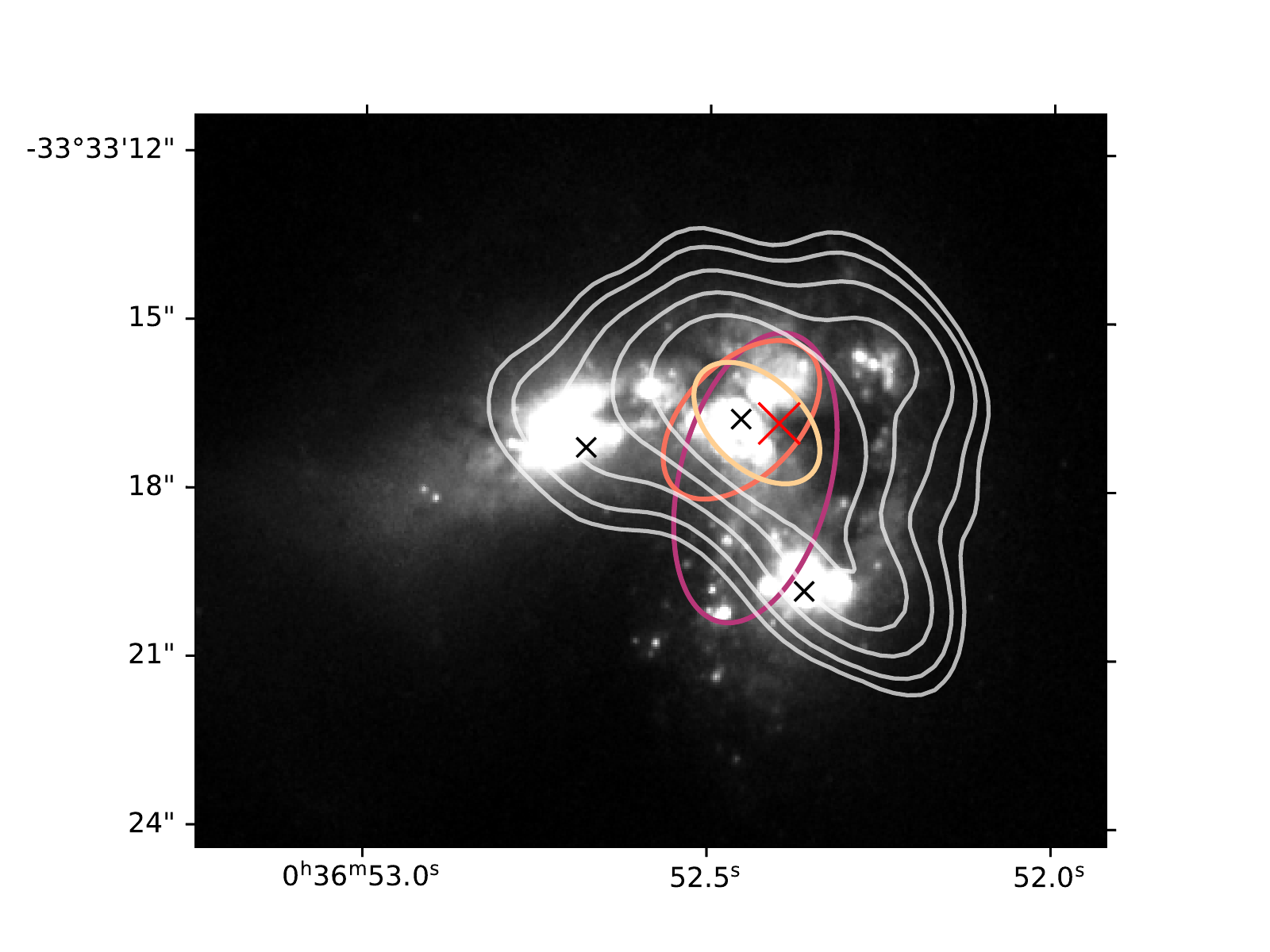}
    \caption{Radio continuum and optical stellar emission comparison. HST F435W filter image with VLA continuum emission contours overlaid in pink for the S-band (3GHz), in orange for the X-band (9.8GHz), in yellow and white for the Ka-band (33GHz). The contours for each band are also presented on Fig. \ref{fig:radio_cont}. The colorful contours correspond to the Gaussian aperture used to extract the flux in Knot B, while the white contours show the Ka-band continuum at the \{3,5,10,20,40\}$\times \sigma$ levels. Black crosses indicate the position of the three star-forming knots, the red cross indicates the position of the 21cm absorption centroid.}
    \label{fig:Haro11_abs_centr}
\end{figure}

On Fig.~\ref{fig:Haro11_abs_centr} we present a comparison of radio continuum and stellar emission. Radio continuum emission contours and the 21cm absorption centroid are overlaid on an HST F435W filter image of the stellar emission in Haro 11. Colorful contours highlight the location of the unresolved continuum source in Knot B, while white contour trace the morphology of the radio continuum emission with the highest resolution radio continuum image.

\newpage
\section{21cm morpho-kinematics}
\begin{figure*}
    \centering
    \includegraphics[width=\textwidth]{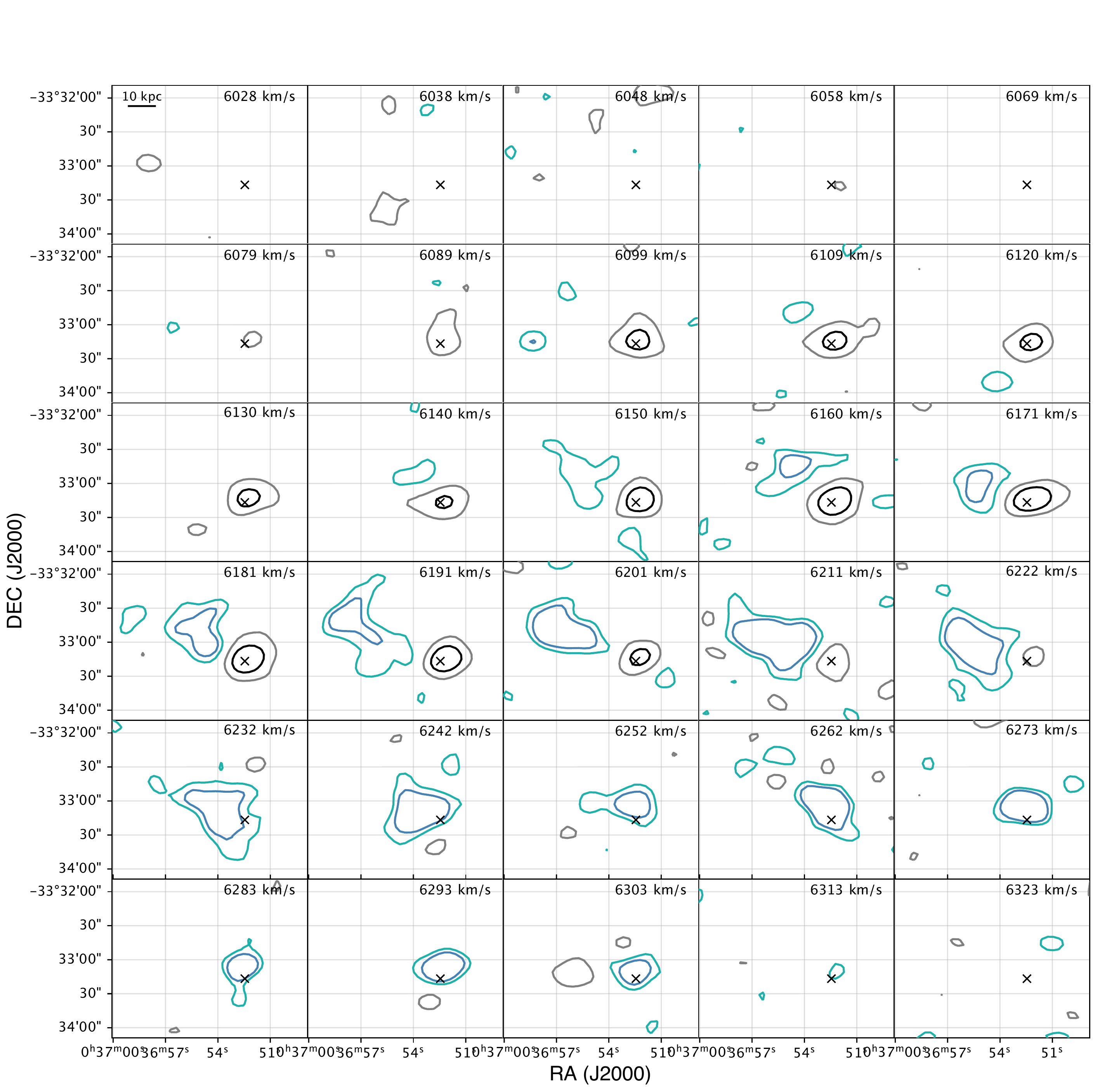}
    \caption{MeerKAT intermediate angular resolution 21cm channel map. Levels displayed correspond to the -10$\sigma$, -3$\sigma$, 3$\sigma$ and 5$\sigma$ levels, respectively shown in black, gray, light blue and dark blue. The position of Knot B is indicated by a black cross. The two other knots are extremely close to Knot B, they have not been represented here to help readability.}
    \label{fig:chanmap_interm_res}
\end{figure*}

\begin{figure*}
    \centering
    \includegraphics[width=\textwidth]{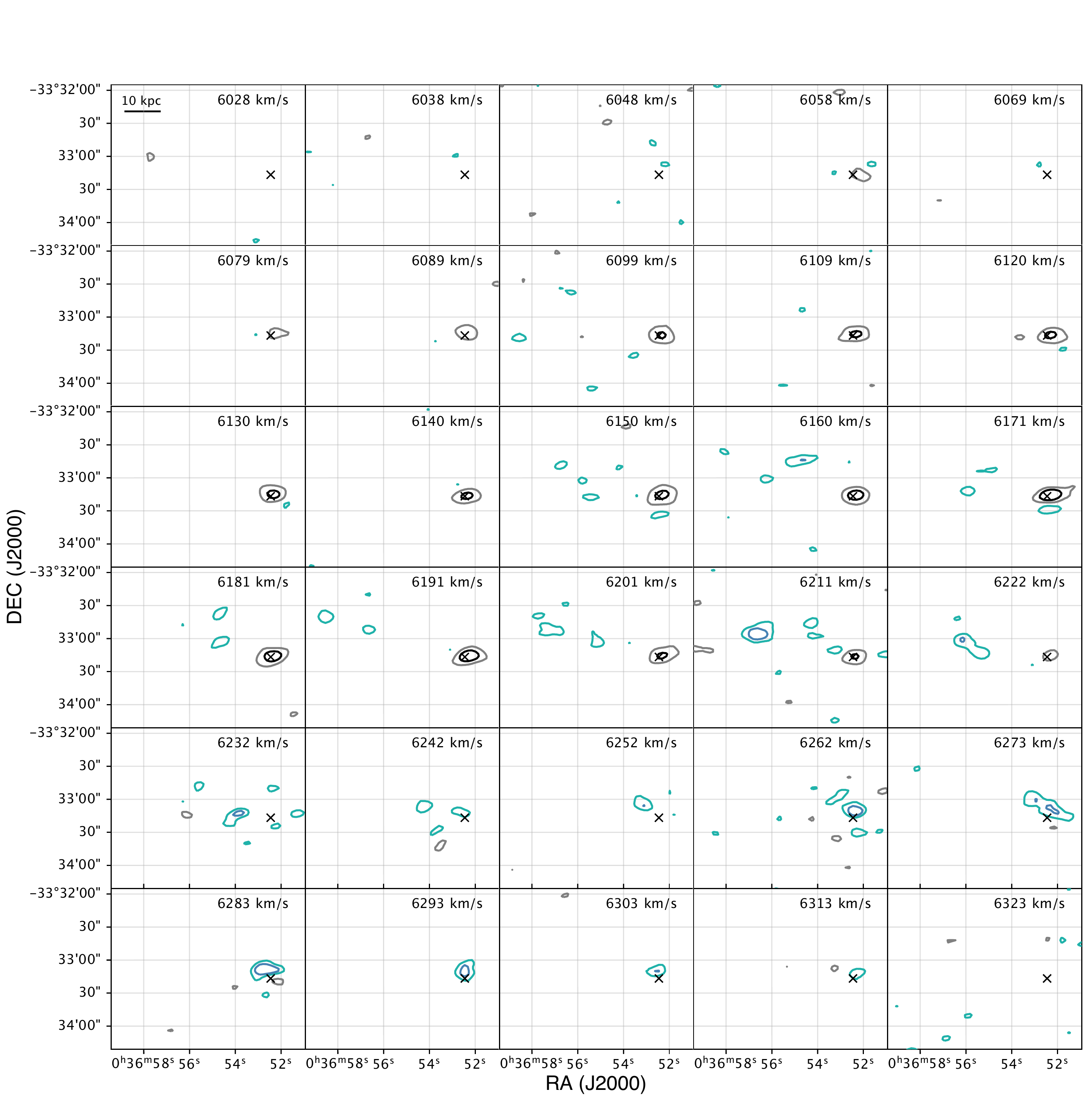}
    \caption{MeerKAT high angular resolution 21cm channel map. Levels displayed correspond to the -10$\sigma$, -3$\sigma$, 3$\sigma$ and 5$\sigma$ levels, respectively shown in black, gray, light blue and dark blue. The position of Knot B is indicated by a black cross. The two other knots are extremely close to Knot B, they have not been represented here to help readability.}
    \label{fig:chanmap_high_res}
\end{figure*}

Channel maps of the 21cm of Haro 11 are shown for the high-resolution and intermediate-resolution cubes in Fig. \ref{fig:chanmap_high_res} and \ref{fig:chanmap_interm_res}. A large fraction of the 21cm emission is found in an extended feature that is connected both
spatially and spectrally and is offset from the center of the galaxy. This morphology, along with previous observations concluding to the merger nature of Haro 11 \citep{Ostlin2015}, and the reproduction of the neutral gas structure with simulations,  strongly suggest that the extended 21cm emission traces a tidal tail resulting from the merger interaction.
\newpage          

\newpage

\section{Ionized gas maps}
\begin{figure}
    \centering
    \includegraphics[width=0.47\textwidth]{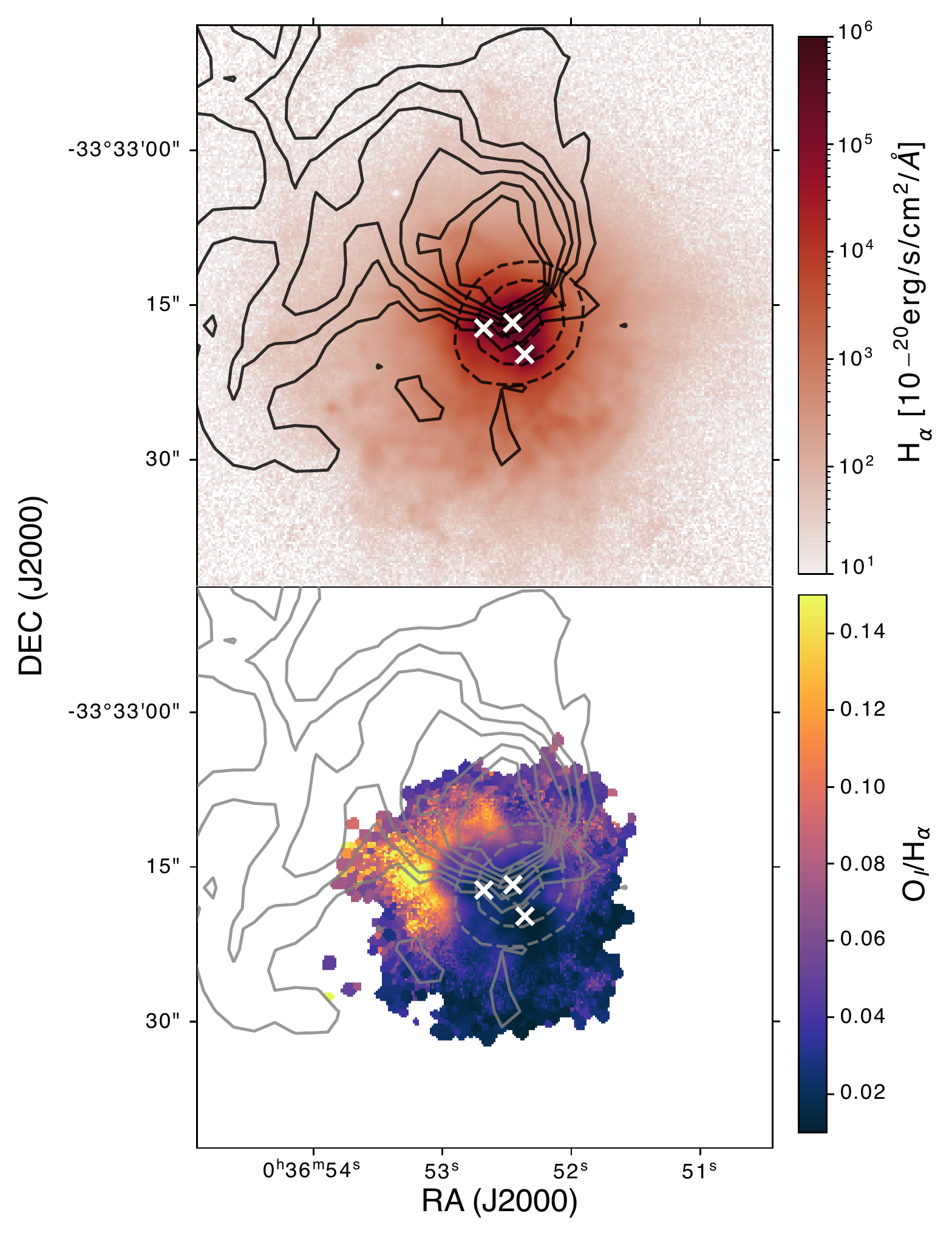}
    \caption{MUSE ionized gas maps with \hi\ contours overlaid. The top panel shows the \ha\ moment-0 map, while the bottom panel shows the \oi/\ha\ ratio map. The \hi\ contours shown in black and grey in the two panels are the same as in Fig. \ref{fig:21cm_global}. The white crosses indicate the positions of the star-forming knots. North is up and East is to the left.}
    \label{fig:MUSE_maps}
\end{figure}
Ionized gas maps from MUSE with neutral gas contours are presented in Fig. \ref{fig:MUSE_maps}. The ionized gas is distributed in a large halo surrounding the knots as seen in the \ha\ line map on the top panel. The \oi/\ha\ line ratio map, shown on the bottom panel, traces shocks, which can be identified by elevated value of the ratio. Interestingly,  elevated values of the ratio (\oi/\ha$>0.1$ ) tracing a highly shocked feature are seen on the North-Eastern part of the Haro 11, towards high \hi\ column density regions. This occurs in regions where the extended neutral gas structure reconnects with the optical body of the galaxy. The most promising explanation for this feature is one where the cold dense gas provides a working surface to shock hot, outflowing material. This can be explained by the existence of an outflow around the optical body of the galaxy, that is shocking the gas found in the neutral tidal tail.

\section{A case where the absorbing neutral gas mass is concentrated in front of Knot B}
\label{sec:mass_abs_kb}
Here we consider a case where the neutral gas seen in absorption would be concentrated in front of the radio continuum source overlapping with Knot B. We follow the same method outlined in section \ref{sec:HI-mass}, but adjust the parameters used to calculate the mass. We use the gas covering fraction $f=0.96$ derived for Knot B \citep{Ostlin2021} and use the upper limit to the size of the radio continuum source derived with Ka-band imaging (1.66" $\times$2.71"). We approximate the continuum flux in Knot B at 1.4GHz as 25 mJy by extrapolating the radio continuum SED fit. These parameters yield an absorbing gas mass $\textrm{M}_{\rm{HI,abs}} = 2.8\pm1.1 \times 10^{7} \,\textrm{M}_{\odot}$. However, due to angular resolution limits of our observation, we cannot determine if the absorbing gas is in fact fully distributed in front of this knot. UV absorption line observations of the knots identify neutral gas outflows in front of other parts of the galaxy, indicating that Knot B is not the only region covered by neutral gas in the galaxy \citep{Sirressi2022}. Nevertheless, we note that if the hypothesis of neutral gas fully covering Knot B held true, and no additional gas was observed in emission around the other knots, it would increase both the fraction of mass seen in emission  ($93.2^{+2.9}_{-3.5}\%$), and the associated neutral gas centroid offset ($\sim15\,\textrm{kpc}$).

\section{21cm emission centroid in Haro 11 and THINGS galaxies}
\begin{figure*}
    \centering
    \includegraphics[width=\textwidth]{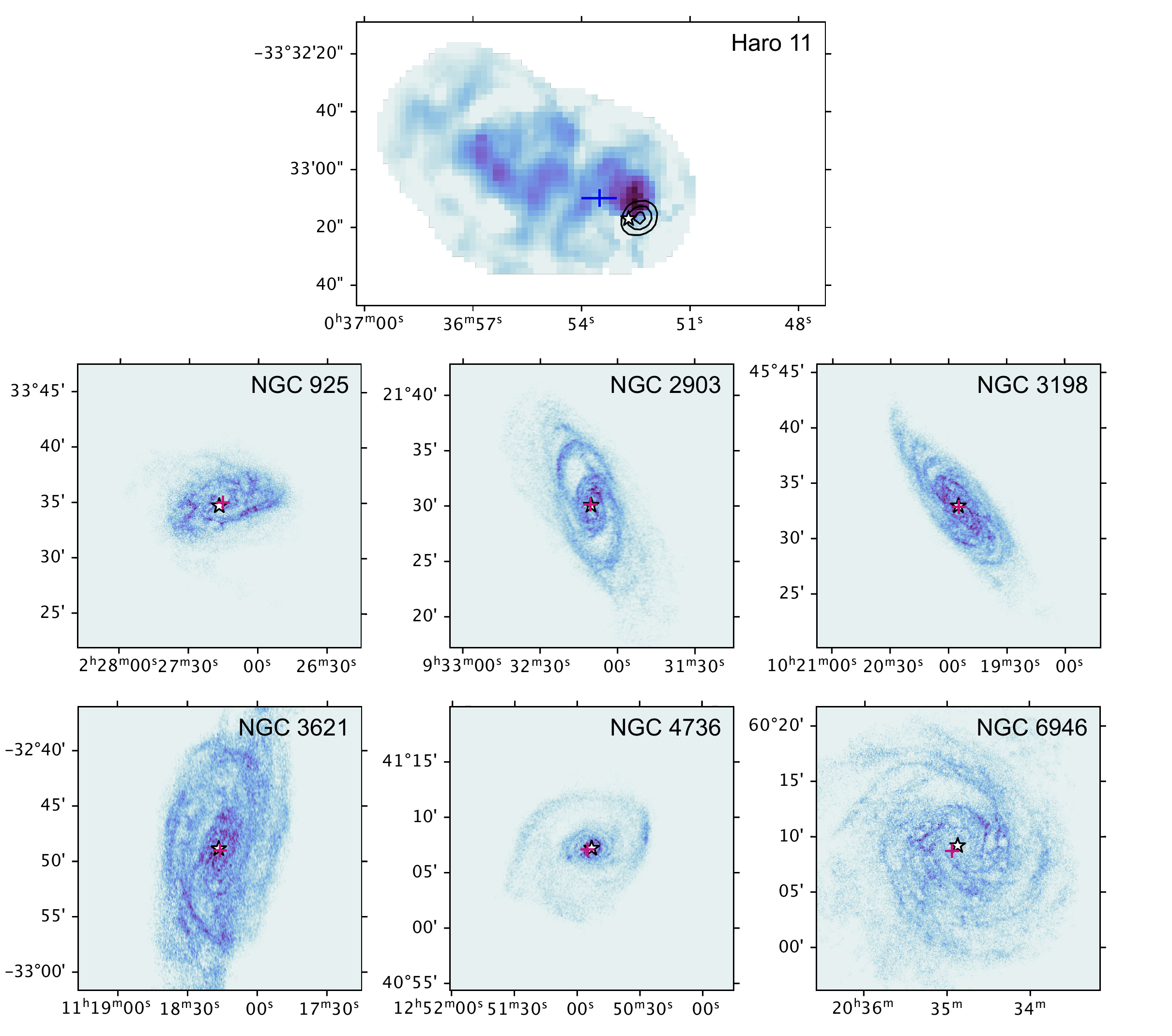}
    \caption{Position of 21cm centroids compared to the optical coordinates in Haro 11 and THINGS galaxies of similar stellar mass, overlaid on 21cm emission images. In each image, the optical coordinates of the galaxies are indicated by a white star. The 21cm centroid of Haro 11 (top panel) is indicated by blue error bars taking into account the uncertainty in absorbing gas mass. We show the same contours of absorbing as in Fig. \ref{fig:21cm_global}. The 21cm emission centroids of THINGS galaxies are shown by purple crosses.}
    \label{fig:21cm_centroid_images}
\end{figure*}

We show the position of 21cm centroids of Haro 11 and THINGS galaxies of similar stellar mass on Fig. \ref{fig:21cm_centroid_images}. The centroids, shown by purple or blue crosses, are overlaid on 21cm emission images, and are compared to the optical coordinates indicated by white stars. For Haro 11, we have also overlaid the 21cm absorption contours, which are taken into account when calculating the 21cm centroid.


\bsp	
\label{lastpage}
\end{document}